\documentclass{article}

\usepackage[main, final]{neurips_2025}

\usepackage[utf8]{inputenc} 
\usepackage[T1]{fontenc}    
\usepackage{hyperref}       
\usepackage{url}            
\usepackage{booktabs}       
\usepackage{amsfonts}       
\usepackage{nicefrac}       
\usepackage{microtype}      
\usepackage{xcolor}         
\usepackage[table]{xcolor}
\usepackage{url}
\usepackage[utf8]{inputenc}

\usepackage{amsmath}
\usepackage{amssymb}
\usepackage{multirow}
\usepackage{booktabs}

\usepackage[utf8]{inputenc}
\usepackage[most]{tcolorbox} 
\usepackage{cleveref}

\usepackage{fontawesome5} 

\usepackage{fancyhdr}

\usepackage{newtxtext,newtxmath} 
\usepackage[most]{tcolorbox}
\usepackage{xcolor}
\usepackage{subcaption}
\definecolor{BoxBg}{RGB}{253, 253, 253}
\definecolor{TitleBg}{RGB}{245, 245, 245}
\definecolor{FrameColor}{RGB}{120, 120, 120}

\newtcolorbox{promptbox}[1]{
    enhanced,
    colback=BoxBg,
    colframe=FrameColor,
    boxrule=0.5pt,
    sharp corners,
    top=4mm, bottom=3mm, left=4mm, right=4mm,
    fonttitle=\normalfont,        
    coltitle=black,
    colbacktitle=TitleBg,
    title={#1},
    attach boxed title to top left={yshift=-2.5mm, xshift=3mm},
    boxed title style={
        boxrule=0.5pt,
        colframe=FrameColor,
        sharp corners,
    }
}

\newcommand{\token}[1]{{\texttt{\small <|#1|>}}}

\title{Pailitao-VL: Unified Embedding and Reranker for Real-Time Multi-Modal Industrial Search}

\author{
    \begin{tabular}{c}
        Lei Chen, \; Chen Ju\textsuperscript{\faEnvelope}, \; Xu Chen, \; Zhicheng Wang, \\
        Yuheng Jiao, \; Hongfeng Zhan, \;  Zhaoyang Li, \; Shihao Xu, \;
        Zhixiang Zhao, \\ Tong Jia, \; 
        Lin Li, \; Yuan Gao, \; Jun Song, \; Jinsong Lan, \; Xiaoyong Zhu, \; Bo Zheng \\
        \quad \\
        Alibaba Group \\
        {\small \texttt{\{realchenleicu,cju.void\}@gmail.com, huaisong.cx@alibaba-inc.com}}
    \end{tabular}
}

\begin{document}
\maketitle

\begin{abstract}
This paper introduces Pailitao-VL, a unified multi-modal retrieval system specifically engineered for the high-precision, real-time industrial search. We here address three critical challenges in the current SOTA solution: insufficient retrieval granularity, vulnerability to environmental noise, and prohibitive efficiency-performance gap.

\vspace{0.1cm}
\noindent \underline{\textit{Pailitao-VL-Embedding}}: Despite significant advances in embedding, a substantial gap persists between coarse concept-level alignment and instance-level precision required for the industrial search. Current solutions, predominantly optimized via contrastive learning on web-crawled data, excel at distinguishing broad categories, such as a sedan from an SUV, but often fail to resolve subtle intra-concept variations, {\em e.g.}, identifying a specific facelifted car model by its minor headlight contours. This limitation stems from the inherent stochasticity and ``semantic drift'' of contrastive objectives in billion-scale environments, where reliance on local, relative proximity fails to establish the absolute identity necessary for high-fidelity retrieval. To bridge the gap, we explore two synergistic breakthroughs. We first transition to absolute ID-recognition, replacing open-ended alignment with deterministic supervision. By anchoring instances to billion-scale semantic prototypes, we reformulate objective as ultra-scale classification, enabling the backbone to resolve fine-grained structural differences required for instance-level discrimination. To ensure the label purity essential for an absolute space, we develop agent-driven high-capacity data curation, {\em i.e.}, employing one ``plan-propose-organize-review'' pipeline to transform noisy industrial manifolds into high-purity ID clusters, establishing a reliable supervisory that far exceeds the capabilities of traditional heuristic mining.

\vspace{0.1cm}
\noindent \underline{\textit{Pailitao-VL-Reranking}}: 
While high-precision embeddings effectively narrow the search space, generative reranking often constitutes the primary bottleneck, struggling to reconcile fine-grained retrieval with the operational efficiency. Existing pointwise solutions evaluate candidates in isolation, suffering from binary reductionism, where complex relevance is collapsed into a coarse ``Yes/No'' judgment, and a total lack of comparative context between candidate documents. To transcend these constraints, we evolve the paradigm from pointwise to one ``compare-and-calibrate'' listwise policy. By synergizing chunk-based comparative reasoning with calibrated absolute relevance scoring, our system captures nuanced relative preferences within a shared context while ensuring global consistency across independently processed chunks. Such one hybrid ranking policy facilitates massive parallel processing, effectively neutralizing prohibitive computational overhead of large-scale MLLM inference and achieving great speedup in throughput, thereby bridging the gap between peak ranking precision and real-time industrial adoptability.

\vspace{0.1cm}
Extensive evaluations confirm that Pailitao-VL achieves state-of-the-art performance while meeting strict industrial latency requirements. The online A/B tests on the Alibaba’s e-commerce platform demonstrate substantial business value, such as a 20\% GMV gain in some specific AI search scenarios.
\end{abstract}

\vspace{0.7cm}
\tableofcontents 
\newpage         

\section{Introduction} \label{intro}
The exponential proliferation of multi-modal content across the internet has fundamentally reshaped the landscape of digital information. Modern ecosystems are increasingly defined by a heterogeneous mix of data modalities, ranging from natural images and text to complex infographics and videos, necessitating retrieval systems that transcend traditional text-only paradigms. In demanding industrial contexts such as e-commerce and intellectual property protection, the objective of multi-modal search has evolved: it is no longer sufficient to achieve broad concept-level alignment ({\em e.g.}, identifying a category). Instead, systems must now deliver instance-level discriminability, capturing minute structural and textural differences to resolve subtle intra-concept variations. This shift makes high-precision multi-modal retrieval essential for real-world applications, including social media navigation, SKU product discovery, and scientific exploration.

To achieve the goal, contemporary multi-modal retrieval systems typically adopt a cascaded framework consisting of two sequential stages: coarse-grained embedding followed by fine-grained reranking. This pipeline is built upon the principle of progressive refinement, which systematically narrows down the candidate pool to converge on most relevant results while balancing computational throughput with semantic depth. Specifically, the embedding stage enables rapid, large-scale candidate generation across the entire search space by projecting heterogeneous data into a unified latent space. Subsequently, the reranking stage performs exhaustive cross-modal semantic reasoning over a significantly smaller candidate set, delivering high-precision alignment necessary for complex, real-world queries.

The primary objective of embedding is to reliably retrieve one hundred-scale candidate set from billion-scale corpora. The standard solution involves projecting heterogeneous image-text data into a unified, low-dimensional latent space where semantic similarity is evaluated by vector proximity, {\em e.g.}, cosine similarity. Optimization is typically driven by large-scale contrastive alignment, which constructs a shared manifold by pulling positive pairs together while pushing negatives apart. Architecturally, the field has evolved from traditional CLIP-style dual-encoders~\cite{radford2021learning} to contemporary MLLM2vec-style architectures~\cite{jiang2024vlm2vec, jiang2024e5-v, zhang2024gme} that leverage sophisticated attention dependencies for deeper alignment. For deployment, these embeddings facilitate the offline pre-computation of billion-scale repositories, which are indexed in the specialized vector databases ({\em e.g.}, Faiss or Milvus) to achieve sub-millisecond retrieval latency. However, condensing the rich multi-modal inputs into a single fixed-dimensional vector inherently creates an information bottleneck, fundamentally limiting the system's capacity for fine-grained discrimination.

The primary objective of reranking is to perform refinement for the candidate results within a drastically narrowed search space, typically operates on a constrained set of hundred-scale documents per query. The reranking architecture has evolved from early cross-encoders of shallow Transformer layers, to contemporary MLLMs~\cite{sun2023chatgpt, liu2025lamra}, facilitating exhaustive semantic interaction. By taking query-document pairs as input, these models are often trained to generate designated tokens ({\em e.g.}, ``Relevant/Irrelevant''), using their predicted probabilities as calibrated reranking scores. Despite superior precision, the massive computational overhead of deep cross-modal reasoning remains a critical barrier to real-time, large-scale industrial deployment.

Despite the significant progress of this two-stage framework, we identify three fundamental challenges that hinder its effective deployment in demanding industrial scenarios:

$\bullet$ \textbf{Insufficient Retrieval Granularity}. 
Most existing solutions optimize for concept- or category-level retrieval, learning coarse alignments from large-scale web-crawled data via contrastive objectives. While effective at distinguishing broad concepts (\emph{e.g.}, sedan \emph{vs.}\ SUV), they lack the instance-level discriminability needed to resolve subtle intra-concept variations (\emph{e.g.}, distinguishing a facelifted model from its predecessor by minor headlight contours), making them fall short in high-precision scenarios demanding identification of specific product versions or minute structural differences.

$\bullet$ \textbf{Pronounced Vulnerability to Domain Shift or Environmental Noise}. 
In production environments, query-document images often deviate substantially from clean, curated pre-training data. Adverse factors, including poor lighting, cluttered backgrounds, physical occlusions, and visual artifacts (OCR overlays, watermarks), induce severe retrieval drift. Existing solutions struggle to disentangle core semantics from such perturbations, leading to degraded performance in unconstrained settings.

$\bullet$ \textbf{Fundamental Efficiency-Performance Gap Preventing Real-Time Adoption}. 
Recent performance gains often come at the cost of massive parameterization, complex cross-modal interactions, and autoregressive decoding, all of which impose prohibitive computational overhead. 
For industrial platforms requiring high-concurrency processing (thousands of queries per second), bridging the gap between peak semantic precision and operational throughput remains a primary bottleneck.

To bridge the gaps, we present Pailitao-VL, a comprehensive suite comprising Pailitao-VL-Embedding and Pailitao-VL-Reranker, specifically engineered for high-precision, real-time industrial search. Built upon the foundation of Multi-Modal Large Language Model (MLLM), our framework integrates advanced data governance, efficient architecture with specialized large-scale training methodologies.

Pailitao-VL-Embedding departs from traditional contrastive learning by introducing semantic categorization for absolute ID-identification. Its central is an MLLM-driven data governance agent that utilizes a ``Plan-Propose-Organize-Review'' pipeline to purify raw industrial corpora into a billion-scale semantic prototype library. By reformulating retrieval as ultra-scale classification, we map instances to these prototypes as absolute semantic anchors in the latent space. Optimized via an additive margin angular objective in the three-stage pipeline, the embedding resolves minute structural and textural nuances, achieving the high-precision discriminability required for instance-level identification.

Complementing the embedding, the Pailitao-VL-Reranker series evolves the reranking paradigm from isolated pointwise evaluation to a sophisticated ``compare-and-calibrate'' listwise process. To resolve the efficiency-performance trade-off, we introduce a hybrid ranking policy that synergizes chunkwise local ranking with absolute relevance scoring. While chunkwise reasoning allows the model to capture fine-grained relative preferences across candidate groups, absolute relevance scoring, grounded in a fixed four-level hierarchy, ensures global consistency across independently processed chunks. This architecture facilitates parallelized processing, significantly reducing end-to-end latency while maintaining peak ranking precision.

We evaluate the Pailitao-VL series across one rigorous suite of offline benchmarks and large-scale industrial deployments. Experimental results demonstrate that our models achieve state-of-the-art performance offline, while extensive online A/B testing on the Alibaba e-commerce platform confirms their substantial business vlaue and operational efficiency. Concretely, Pailitao-VL-Embedding and Pailitao-VL-Reranker-List achieve optimized inference latencies of 67 ms and 76 ms per query, meeting stringent requirements of high-concurrency production environments. Multi-Day average statistics reveal that these technical breakthroughs translate directly into tangible growth: Pailitao-VL-Embedding delivers a 2\% GMV gain across platform-wide traffic, while Pailitao-VL-Reranker-List yields a 6\% GMV increase within standardized product categories. Notably, in emerging AI-driven scenarios such as SKU-price comparison, our architecture achieves an impressive 20\% GMV gain, showing the immense potential of Pailitao-VL for modern, large-scale practical search systems.

\section{Related Work}
\label{relatedwork}
\noindent\textbf{Multi-Modal Embeddings.}
Early approaches to multi-modal information retrieval were predominantly characterized by dual-encoder architectures, exemplified by models such as CLIP~\citep{clip}, ALIGN~\citep{align}, and SigLIP~\citep{siglip}. These baselines leverage contrastive learning to map images and text into a shared embedding space, where representational similarity is designed to correlate with semantic relevance. A fundamental limitation of this paradigm, however, is its inherent difficulty in handling complex, interleaved visual and linguistic inputs, as the separate encoders preclude deep, early-stage fusion of modalities.

To address this challenge, subsequent methods explored for richer multi-modal integration. Initial efforts, UniVL-DR~\citep{univl-dr} and UniIR~\citep{wei2024uniir}, adopted late-stage fusion strategies to combine features extracted from separate CLIP~\citep{clip} or BLIP~\citep{li2022blip} encoders. Concurrently, VISTA~\citep{zhou2024vista}, Holism~\citep{wang2025advancing} and MARVEL~\citep{zhou2024marvel} augmented powerful pretrained text encoders with visual ``plug-in''.

A more profound paradigm shift has emerged with the repurposing of MLLMs as universal embedding encoders. By leveraging contrastive fine-tuning, pioneering frameworks such as E5-V~\citep{jiang2024e5-v}, VLM2Vec~\citep{jiang2024vlm2vec}, LamRA~\citep{liu2025lamra}, PDF~\cite{wang2025explore}, and GME~\citep{zhang2024gme} have adapted MLLMs to generate high-quality, unified representations capable of distilling interleaved image-text content. To further optimize these foundations, CAFe~\citep{yu2025cafe} synergizes contrastive objectives with autoregressive language modeling losses, aiming to preserve the MLLM's generative versatility while bolstering its discriminative precision. Complementing these efforts, LLaVE~\citep{lan2025llave} and UniME~\citep{gu2025unime} focus on refining embedding quality through sophisticated hard-negative mining. Pushing the frontier even further, TTE~\citep{cui2025tte} and UME-R1~\citep{lan2025ume-r1} have introduced the concept of ``generative embeddings''. In this paradigm, the model first produces explicit Chain-of-Thought reasoning~\citep{wei2022cot} or semantic summaries prior to embedding generation, effectively infusing the final latent vector with the model's internal reasoning process to achieve deeper semantic density.

In all aforementioned methods, retrieval is performed by computing embedding similarity between query and document. While powerful, these methods, which are predominantly trained via contrastive learning, face two key limitations. First, the resulted embeddings often lack a sufficiently fine-grained understanding of instance-level details. Second, their performance is heavily reliant on the challenging task of mining effective hard negatives. In contrast, we diverge from this embedding learning paradigm by semantic ID-recognition for absolute anchor learning. Besides that, we reformulate retrieval as the task of discriminative classification and re-ranking, thereby directly leveraging the powerful prior knowledge and implicit reasoning capabilities of MLLMs to reorder documents.

\noindent\textbf{LLM for Reranking.}
Large Language Models have been increasingly employed for reranking tasks, which can be broadly categorized into pointwise and listwise strategies.

Pointwise re-rankers independently assess the relevance of individual query-document pairs~\citep{sun2023chatgpt, ma2024fine, liu2025lamra, lin2024mm-embed, niu2024judgerank}. In this paradigm, the model generates a standalone relevance score for each candidate, with the final ranking derived through a subsequent descending sort. While pointwise effectively reframes ranking as a classification or regression task, its optimization objective often remains misaligned with global ranking metrics. Notably, by evaluating documents in isolation, pointwise fails to capture the comparative context among candidates, overlooking the relative semantic nuances essential for high-precision discrimination.

Conversely, listwise re-rankers jointly process all candidates of one query to produce a holistically ordered permutation~\citep{sun2023chatgpt, liu2025lamra, zhuang2024setwise, zhuang2025rank-r1}. The primary strength lies in its capacity for side-by-side comparison, using global contexts to resolve subtle relative nuances that are often imperceptible in isolation. Moreover, the listwise objective allows for training that aligns directly with ranking-specific metrics such as NDCG. However, these gains in precision often come at a prohibitive computational cost. Existing listwise solutions frequently suffer from severe inference latency and memory overhead, as the requirement to ingest exhaustive candidate lists as a single, long input sequence creates a great efficiency bottleneck for real-time deployment.

To mitigate the computational burden of extensive input contexts, we introduce a hybrid architecture that decomposes global listwise reranking into parallelizable chunks. By synergizing chunkwise local ranking with absolute relevance scoring, the system effectively circumvents the latency bottlenecks inherent in processing exhaustive candidate lists. And this design facilitates massively parallelized inference across multiple document chunks, drastically compressing end-to-end response times. Crucially, while the local chunks enable deep comparative reasoning, the integration of calibrated absolute anchors ensures seamless global consistency across chunk boundaries, striking an optimal balance between semantic depth and operational efficiency.

\section{Task Formalization} \label{method} 
Let $\mathcal{D} = \{\mathrm{d}_i\}_{i=1}^L$ denote one large-scale corpus consisting of $L$ documents ($L \approx 10^9$), where each document $\mathrm{d} = (\mathrm{d}_\mathrm{img}, \mathrm{d}_\mathrm{txt})$ consisting of image-text content. For a multi-modal query $\mathrm{q} = (\mathrm{q}_\mathrm{img}, \mathrm{q}_\mathrm{txt})$ given by the user, our objective is to retrieve and rank documents from $\mathcal{D}$ that best meets the relevance criteria. Note that $\mathrm{q}_\mathrm{txt}$ is usually a high-precision requirement for instance-level discrimination of subtle intra-concept variations, with some spelling errors, informal abbreviations, {\em etc}. While $\mathrm{q}_\mathrm{img}$ is often noise-aware, {\em i.e.}, the image may be subject to low lighting, cluttered backgrounds, physical occlusions, low resolution and ``visual noise'' (OCR overlays or advertising watermarks).

Following the modern multi-modal retrieval paradigm, we formulate the task as: \textit{coarse-grained discriminative embedding} followed by \textit{fine-grained generative reranking}.

\textbf{Coarse-Grained Discriminative Embedding} $\Phi_\mathrm{embd}$ aims to map the query and documents into a visual-textual alignment latent space $\mathbb{R}^D$, where $D$ refers to the embedding dimension. We retrieve a candidate set $\mathcal{D}_\mathrm{cand} \subset \mathcal{D}$ of $N$ documents ($N \ll L$, typically $N \approx 10^2$) by calculating the embedding similarity between query and documents, and retaining the top-N most similar documents.
\begin{equation}
\mathcal{D}_{\mathrm{cand}} = \mathop{\text{arg-top-}N}_{\mathrm{d}_i \in \, \mathcal{D}} \left( \text{sim} \left( \Phi_{\mathrm{embd}}(\mathrm{q}), \Phi_{\mathrm{embd}}(\mathrm{d}_i) \right) \right), 
\end{equation}
where $\text{sim}(\cdot, \cdot)$ denotes the proximity metric such as cosine similarity. The embedding emphasizes high-throughput recall across the exhaustive search space via Approximate Nearest Neighbor search.

\vspace{0.1cm}
\textbf{Fine-Grained Generative Reranker} $\Phi_{\mathrm{rank}}$ aims to refine relevance results in $\mathcal{D}_\mathrm{cand} = \{\mathrm{d}_1, \dots, \mathrm{d}_N\}$ by capturing intricate, token-level cross-modal interactions. It usually takes the concatenation of query $\mathrm{q}$ and each document $\mathrm{d}_i \in \mathcal{D}_{\mathrm{cand}}$ as input to facilitate exhaustive reasoning. Given the MLLMs' generative nature, we formulate the calibrated scoring as a conditional probability estimation:
\begin{equation}
\mathrm{s}_i = P( \text{``Rel''} \mid \mathrm{q}, \mathrm{d}_i; \Phi_{\mathrm{rank}} ), \quad \forall \mathrm{d}_i \in \mathcal{D}_{\mathrm{cand}},
\end{equation}
where $\mathrm{s}_i \in [0, 1]$ is derived from the logit of a specific positive token ``Rel''. After this, we determine an optimal permutation of the indices $\{1, \dots, N\}$ that reorders $\mathcal{D}_{\mathrm{cand}}$ through $\mathrm{s}$. By resolving the instance-level discrimination and environmental noise that were collapsed in the embedding space, $\Phi_{\mathrm{rank}}$ effectively aligns the final ranking with the user's high-precision intent.

\section{Discriminative Embedding}
The embedding $\Phi_{\mathrm{embd}}$ serves as the cornerstone of multi-modal retrieval. In this section, we begin by reviewing existing approaches and subsequently present our solution: semantic ID-recognition for absolute anchor learning (termed as \textbf{Pailitao-VL-Embedding}).

\subsection{Contrastive Learning Paradigm}
Despite significant progress in the embedding architecture (from separated dual-encoders to VLM2vec), the training paradigm still follows the old path of contrastive learning. Typically, mainstream methods construct positive-negative pairs based on the self-supervised signals, such as image-text co-occurrence in general web data, implicit user behavior in e-commerce platforms ({\em e.g.}, query-click/non-click logs) or consistency across multi-view images of the same ID product. These models are primarily optimized via contrastive objectives like InfoNCE, aiming to pull positive pairs together while pushing apart randomly sampled in-batch negatives or mined hard negatives.

However, this contrastive paradigm inherently struggles with high-precision, instance-level retrieval, which motivates us to advocate for one shift toward an absolute ID-recognition paradigm. We identify three critical deficiencies of contrastive learning that a ID-recognition idea can effectively address:

\underline{\textit{Stochasticity vs. Deterministic Supervision}}: Contrastive learning relies heavily on quality of negative pairs. Random in-batch negatives are usually too semantically distant to provide effective gradients, while heuristic hard-negative mining frequently introduces ``false negatives'' ({\em e.g.}, pulling apart different views of the same ID). This stochastic supervision leads to unstable decision boundaries. The ID-recognition paradigm eliminates this noise by replacing random pair-mining with deterministic labels, ensuring that every gradient update is guided by high-quality, pre-defined instance-ID clusters.

\underline{\textit{Relative Proximity vs. Absolute Identity}}: The contrastive objective only optimizes for relative distances within a local batch, rather than anchoring instances to a global semantic coordinate. Consequently, the latent space lacks global consistency, and embeddings tend to ``drift'' across training stages. By reformulating the task as ultra-scale ID-recognition, each instance is provided with an absolute semantic identity, ensuring that every document is aligned with a stable, globally-consistent prototype, acting as a permanent anchor in the latent space.

\underline{\textit{Coarse-Grained Partitioning vs. High-Precision Resolution}}: Standard contrastive methods lack an explicit mechanism to partition the latent space into high-resolution regions, often leading to coarse embeddings. An ultra-scale ID-recognition addresses this by utilizing a global identification head to supervise exhaustive clusters. By strategically sampling proximal ID centers, it imposes a much more rigorous discriminative constraint, compelling the backbone to capture minute structural or textural differences that are typically ignored by coarse-grained contrastive objectives.

\begin{figure}[t]
\begin{center}
\includegraphics[width=1\textwidth] {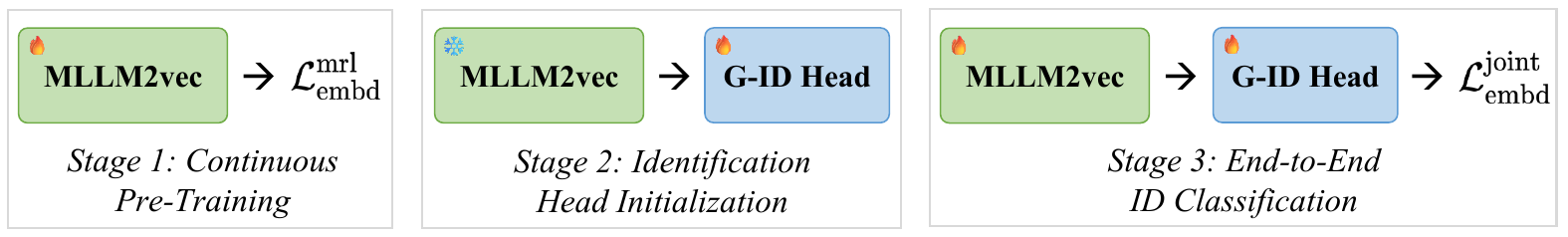}
\end{center}
\vspace{-0.3cm}
\caption{\textbf{Pailitao-VL-Embedding} evolves from popular contrastive learning to absolute prototype-ID recognition, enhancing instance-level discriminative capability through three-stage optimization.}
\label{fig:intro}
\end{figure}

\subsection{Absolute ID-Recognition Paradigm}
Our Pailitao-VL-Embedding constructs a fundamental paradigm shift in both data governance and optimization objective, transitioning from stochastic pair-based contrast to deterministic ID-based discrimination. Next, we introduce the three-stage training pipeline: firstly, continuous pre-training with image-text contrastive learning on the MLLM2vec backbone; secondly, bootstrapping a global identification head initialized by billion-scale ID prototypes to establish absolute semantic anchors; and finally, conducting end-to-end joint optimization of the MLLM2vec backbone and the identification head to identify the instance-IDs through ultra-scale deterministic classification.

\subsubsection{Stage 1: MLLM2vec $\Phi_{\mathrm{embd}}^{\mathrm{mllm}}$} 
Recognizing the superior cross-modal potential of MLLMs over traditional dual-encoder architectures, we build the backbone following the VLM2vec framework~\cite{jiang2024e5-v,jiang2024vlm2vec,liu2025lamra,zhang2024gme}, by employing a 3B-parameter dense model as the initialization.

\begin{promptbox}{Prompt Template for MLLM2vec Embedding $\Phi_{\mathrm{embd}}^{\mathrm{mllm}}$}
    
    \token{im\_start} system \\
    You are a relevance embedding generator. Given a query or a document, each consisting of an image and a text description, please summarize its semantic embedding.  \\
    \token{im\_end}
    
    \medskip
    
    \token{im\_start} user \\
    Input Text: \{\small$\mathrm{x}_\mathrm{txt}$\} \\
    Input Image: \texttt{<image: $\mathrm{x}_\mathrm{img}$>} \\
    \token{im\_end}
    
    \medskip
    
    \token{im\_start} assistant \\
    \token{EMB}
\end{promptbox}

\vspace{0.1cm}
\noindent \underline{\textit{Input/Output Formulation}}.
Given one multi-modal data $\mathrm{x} \in \{\mathrm{q}, \mathrm{d}\}$, where query $\mathrm{q} = (\mathrm{q}_\mathrm{img}, \mathrm{q}_\mathrm{txt})$ and document $\mathrm{d} = (\mathrm{d}_\mathrm{img}, \mathrm{d}_\mathrm{txt})$, the MLLM2vec backbone $\Phi_{\mathrm{embd}}^{\mathrm{mllm}}$ is employed to project $\mathrm{x}$ into a unified latent space. Specifically, as shown in the above prompt template, we construct an input sequence $\mathcal{E}$ by prepending a task-specific instruction $\mathcal{T}_{\mathrm{inst}}$ to the multi-modal content:
\begin{equation}
\mathcal{E} = \texttt{Concat}\!\left(\, \mathcal{T}_{\mathrm{inst}},\; \mathrm{x}_\mathrm{txt},\; \mathrm{x}_\mathrm{img} \,\right),  \quad  \mathrm{x} \in \{\mathrm{q}, \mathrm{d}\}.
\end{equation}

Following the decoder-only paradigm, we append a special \token{EMB} token to the end of the input sequence, allowing it to aggregate contextual information from all preceding tokens through causal attention. We then extract the final hidden state $\mathbf{h} \in \mathbb{R}^{D}$ at the \token{EMB} position as the contextualized embedding of $\mathrm{x}$. During retrieval, we compute the embedding similarity between the query and all candidate documents, and retain the top-$N$ documents with the highest scores, thereby enabling high-throughput recall of the candidate set.

\vspace{0.1cm}
\noindent \underline{\textit{Continuous Pre-training with MRL}}.
Initialized from general-purpose MLLM checkpoints, we conduct continued pre-training on a massive industrial-scale corpus comprising approximately two billion image-text pairs and curated (anc, pos, neg) triplets. To satisfy diverse downstream requirements, particularly the need for low-cost storage in specialized vector databases, we integrate \textit{Matryoshka Representation Learning} (MRL) into the optimization objective. We define a set of target dimensions $\mathcal{G} = \{ 256,512,1024,2048,3072 \}$, then for each granularity $g \in \mathcal{G}$, the embedding $\mathbf{h}$ is truncated to its first $g$ components to compute one dimension-specific contrastive loss.
Let $\mathbf{h}^{\mathrm{que}} = \Phi_{\mathrm{embd}}^{\mathrm{mllm}}(\mathrm{q})$ and $\mathbf{h}^{\mathrm{doc}} = \Phi_{\mathrm{embd}}^{\mathrm{mllm}}(\mathrm{d})$ be the full-dimensional embeddings for query and document, the total MRL objective is formulated as the weighted sum of InfoNCE losses across all granularities:
\begin{equation}
\mathcal{L}_{\mathrm{embd}}^{\mathrm{mrl}} = \sum_{g \in \mathcal{G}} \lambda_g \cdot \left( -\frac{1}{|\mathcal{B}|} \sum_{i \in \mathcal{B}} \log \frac{\exp(\mathrm{sim}(\mathbf{h}_{i, 1:g}^{\mathrm{que}}, \mathbf{h}^{\mathrm{doc}+}_{i, 1:g}) / \tau)}{\sum\limits_{{j \in \mathcal{B}}} \exp(\mathrm{sim}(\mathbf{h}_{i, 1:g}^{\mathrm{que}}, \mathbf{h}_{j, 1:g}^{\mathrm{doc}}) / \tau)} \right),
\end{equation}
where $\lambda_g$ is the balancing weight for granularity $g$, $\tau$ is the temperature, and $\mathcal{B}$ is the mini-batch. $\mathbf{h}^{\mathrm{doc}+}$ represents the positive document corresponding to query (derived from image-text pairs or triplets), while the denominator incorporates both randomly sampled in-batch negatives and mined hard negatives to provide one rigorous contrastive signal. This multi-granularity optimization objective effectively compels the MLLM backbone to compress the most critical discriminative features into lower-dimensional prefixes of the embedding space. As a result, Pailitao-VL-Embedding allows for flexible trade-offs between retrieval precision and computational efficiency.

Upon completion of this stage, the MLLM2vec backbone exhibits robust concept-level alignment, effectively grouping semantically related images and texts. However, relying solely on contrastive signals derived from noisy co-occurrence or user behavior remains insufficient for instance-level discrimination, as it frequently fails to identify subtle variations required for exact ID matching. This limitation necessitates the subsequent introduction of the global identification head.

\subsubsection{Stage 2: Global Identification Head $\Phi_{\mathrm{embd}}^{\mathrm{head}}$} 
To bridge the gap between relative proximity and absolute identity, we establish a fixed coordinate system for the latent space via the global identification head. And the foundation of this stage is the construction of one high-purity, billion-scale semantic prototype library, where each prototype acts as a deterministic semantic anchor for a unique instance-ID.

\noindent \underline{\textit{Agent-Driven Data Curation}}. 
Recognizing that raw industrial data is inherently plagued by label noise and semantic redundancies, we propose a high-capacity MLLM-based agent system for hierarchical data purification, which operates as a strategic ``Plan-Propose-Organize-Review'' pipeline:

\textbf{Proposers (Multi-Source Knowledge)}: To ensure exhaustive coverage, the MLLM planner orchestrates instance candidates from three hierarchical layers of certainty: (1) Deterministic Metadata: leveraging native structural relations, {\em e.g.}, SKU-tree, as the hard-labeled seeds. (2) Probabilistic Behavioral Logs: distilling user click/buy-through trajectories to bridge diverse queries with target documents via the temporal co-occurrence. (3) Generative Semantic Inference: utilizing the planner to synthesize latent links between visually disparate but ID-identical documents, particularly for bridging the gap between cross-domain catalogs.

\textbf{Organizer (Manifold Structuralization)}: To address unannotated documents and the ``cold-start'' problem, the MLLM planner employs SOTA embedding solutions as topological organizers. Rather than simple clustering, the organizer performs density-based manifold expansion to identify long-tail and unannotated documents that reside in the semantic proximity of the proposers' seeds, effectively absorbing isolated data points into cohesive, high-resolution instance clusters.

\textbf{Reviewer (Consensus Adjudication)}: For each proposed prototype cluster, the MLLM planner acts as a judge, invoking existing SOTA rerankers as high-precision reviewers. This stage implements a multi-view consensus mechanism: the reviewer conducts zero-shot attribute verification and cross-perspective consistency checks. Any document that exhibits semantic drift or structural misalignment with the cluster's centroid is pruned. This iterative refinement ensures that each resulting prototype represents a strictly aligned, unique instance-ID entity.

Through this governance, we distill an exhaustive corpus into the purified clusters. Each prototype-ID is computed as the centroid of its respective cluster in the latent space.

\vspace{0.1cm}
\noindent \underline{\textit{Head Initialization}}.
The global identification head is implemented as an ultra-scale linear projection layer. To mitigate convergence instability typically associated with random initialization in massive label spaces, we explicitly populate the layer weights using the pre-computed centroids of purified prototype-IDs. Under this architecture, each row of the weight matrix serves as a dedicated semantic anchor for a specific instance identity. By anchoring these prototypes as permanent semantic targets, we establish a globally consistent reference that transforms traditionally open-ended retrieval into one deterministic, absolute categorization task. This initialization provides the necessary structural stability and semantic alignment for the subsequent optimization.

\subsubsection{Stage 3: End-to-End Joint Optimization} 
By leveraging the absolute semantic anchors established in Stage 2, we here perform joint optimization of the MLLM2vec backbone $\Phi_{\mathrm{embd}}^{\mathrm{mllm}}$ and the global identification head $\Phi_{\mathrm{embd}}^{\mathrm{head}}$. We reformulate retrieval as an ultra-scale deterministic classification task, compelling the model to resolve minute differences between billion-scale instance-IDs.

\vspace{0.1cm}
\noindent \underline{\textit{Learning via Additive Angular Margin}}. 
To enhance intra-ID compactness and inter-ID separation, we incorporate an additive angular margin penalty~\cite{deng2019arcface} between each embedding and its prototype anchor. Specifically, for a data with prototype-ID label $\mathrm{y}_i \in \{1, \dots, U\}$, where $U$ is the ID cardinality, we denote its MLLM2vec-generated embedding as $\mathbf{h}_i \in \mathbb{R}^D$ and the $u$-th weight vector (anchor) of the identification head as $\mathbf{w}_u$. Both $\mathbf{h}_i$ and $\mathbf{w}_u$ are $\ell_2$-normalized to ensure they reside on a unit $D$-dimensional hypersphere. The classification probability for the target identity $\mathrm{y}_i$ is:
\begin{equation}
P(\mathrm{y}_i \mid \mathbf{h}_i) = \frac{\exp\left(\mathrm{sim}(\theta_{i, \mathrm{y}_i} + \mathrm{margin}) / \tau \right)}{\exp\left(\mathrm{sim}(\theta_{i, \mathrm{y}_i} + \mathrm{margin}) / \tau \right) + \sum_{u \neq \mathrm{y}_i} \exp( \mathrm{sim}( \theta_{i, u}) / \tau) },
\end{equation}
where $\tau$ refers to the temperature, $\theta_{i, u} = \arccos(\mathbf{h}_i^\top \mathbf{w}_u)$ denotes the angle between $\mathbf{h}_i$ and $\mathbf{w}_u$.

\textbf{Intra-ID Compactness}: Through adding the $\mathrm{margin}$ to the target angle $\theta_{i, \mathrm{y}_i}$, the model is forced to compress the embedding distribution within one tighter arc around the absolute semantic anchor, effectively resolving intra-prototype variations such as viewpoint and occlusion.

\textbf{Inter-ID Diversity}: The angular penalty increases the difficulty of target ID assignment, thus maximizing the angular separation between proximal ID centers. By strategically sampling semantically similar prototypes in the batch, MLLM2vec is compelled to capture fine structural-textural differences.

To maintain multi-dimensional flexibility, we extend the classification objective to all granularities in $\mathcal{G}$ via MRL, ensuring that each truncated prefix $\mathbf{h}_{1:g}$ possesses independent and robust discriminative power. The total joint loss is formulated as:
\begin{equation}
\mathcal{L}_{\mathrm{embd}}^{\mathrm{joint}} = \sum_{g \in \mathcal{G}} \lambda_g \cdot \left( -\frac{1}{|\mathcal{B}|} \sum_{i \in \mathcal{B}} \log P(\mathrm{y}_i \mid \mathbf{h}_{i, 1:g}) \right).
\end{equation}
This holistic transformation from relative comparison to absolute, margin-enhanced categorization allows Pailitao-VL-Embedding to establish high-resolution semantic boundaries. For production deployment, the identification head is decoupled, leaving only the MLLM2vec backbone for high-throughput real-time embedding generation. By the end of this stage, MLLM2vec transcends coarse concept-level matching, achieving instance-level precision required for industrial retrieval scenarios.

\vspace{0.1cm}
\noindent \underline{\textit{Hybrid Parallel Training}}. 
Scaling global identification head to billion-scale IDs presents three key challenges: massive data throughput, memory footprint of the $U \times D$ weight matrix, and prohibitive inter-GPU communication overhead. We hence implement hybrid parallel training:

\textbf{Data and Model Parallelism}: We employ standard data parallelism for the MLLM2vec backbone, replicating it across all GPUs to handle massive data batches. Conversely, for the identification head, we utilize tensor parallelism. Given that the weight matrix $\mathbf{W} \in \mathbb{R}^{U \times D}$ can exceed the memory capacity of one GPU, we partition the $U$ prototype anchors across the aggregate memory of GPU cluster, ensuring each device only maintains a subset of global semantic space.

\textbf{KNN-based Sparse Similarity}: To mitigate the inter-GPU communication bottleneck during the cross-entropy calculation, we implement a KNN-comparison mechanism. Rather than computing the similarity between an instance $\mathbf{h}_i$ and all $U$ anchors, which would trigger the massive all-to-all communication, we only consider a sparse subset of proximal anchors. For each target ID, we maintain a similarity matrix between prototypes to identify the most relevant negative anchors (10\% of $U$). This sparse approximation focuses the optimization on the most challenging negative boundaries while significantly reducing cross-GPU traffic and accelerating convergence.

\textbf{Dynamic Proxy Update}: The prototype similarity matrix is updated periodically during training to reflect the evolving latent space. This ensures that the KNN-based selection remains topologically accurate, allowing Pailitao-VL-Embedding to maintain ultra-high-resolution discrimination even under the extreme billion-scale constraints.

\subsection{Self-Loop Gain}
The effectiveness of Pailitao-VL-Embedding is anchored in quality and cardinality of the semantic prototype library. Beyond a static training procedure, we observe a self-loop gain: one virtuous cycle between agent-driven data curation and the discriminative capacity of underlying models.

\vspace{0.1cm}
\noindent \underline{\textit{Recursive Refinement Flywheel}}.
In Pailitao-VL-Embedding, the agent system serves as the forge for high-purity data, where embedding and reranker act as the \textit{organizer} and \textit{reviewer} to structuralize the raw industrial corpus. Crucially, as the models undergo the three-stage optimization, their improved discriminative power allows the agent system to perform even more rigorous data governance, identifying subtler near-duplicate redundancies and rectifying deeper label noises. This enhanced data purity, in turn, provides a superior supervisory signal for subsequent iteration of model training. Such recursive refinements ensure Pailitao-VL-Embedding is not merely a one-time optimization but an evolving paradigm that continuously pushes the upper bound of semantic alignment.

\vspace{0.1cm}
\noindent \underline{\textit{Scaling Laws in Prototype Cardinality}}.
We further investigate the efficacy of prototype-ID scale on retrieval reuslts, and empirically observe a scaling law governed by the cardinality of the identity space: as the number of unique prototypes transitions from billion-scale to billion-scale, the model exhibits a steady, power-law-like improvement in instance-level recall and precision. Unlike traditional scaling which mainly focuses on parameter count or dataset volume, Pailitao-VL-Embedding demonstrates that increasing resolution of the identity space ({\em i.e.}, granularity of absolute semantic anchors) is a critical factor for performance scaling in multi-modal retrieval.

\begin{figure}[t]
\begin{center}
\includegraphics[width=0.9\textwidth] {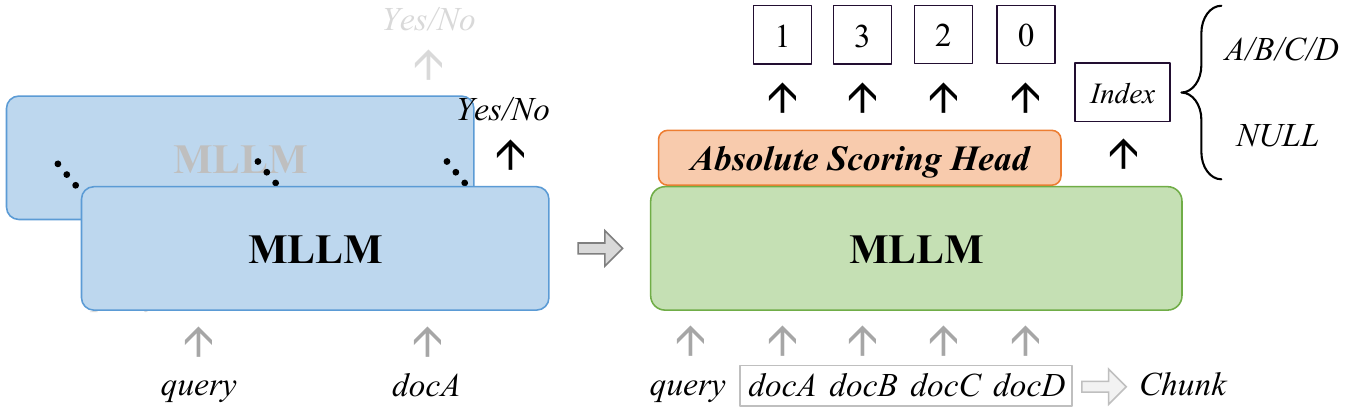}
\end{center}
\vspace{-0.3cm}
\caption{\textbf{Pailitao-VL-Reranker} evolves from pointwise (Pailitao-VL-Reranker-Point) to listwise (Pailitao-VL-Reranker-List), becoming more interpretable, efficient, and performing better.}
\label{fig:intro}
\end{figure}

\section{Generative Reranking} 
While the discriminative $\Phi_{\mathrm{embd}}$ enables high-throughput recall, the inherent semantic compression of its fixed-dimensional embedding naturally constitutes an information bottleneck. To transcend this limitation, we propose the generative \textbf{Pailitao-VL-Reranker} series $\Phi_{\mathrm{rank}}$, which leverages the deep multi-modal comprehension of MLLMs for fine-grained, token-level cross-reasoning. This series encompasses two paradigms: one pointwise reranker for independent candidate evaluation via binary relevance classification, and a listwise reranker for collective comparative reasoning, optimized through multi-level relevance supervision and a hybrid ranking policy.

\subsection{Pointwise Reranker $\Phi_{\mathrm{rank}}^{\mathrm{point}}$}  \label{subsec:pointwise}
Given the hundred-scale document candidate set per query, one natural way is to pair the query with each document, and then input them into MLLM for the binary classification, {\em i.e.}, predicting either a special yes token (relevant) or no token (irrelevant). We refer to this pointwise reranker variant as \textbf{Pailitao-VL-Reranker-Point}.

\noindent \underline{\textit{Input/Output Formulation}}.
Given a candidate document $\mathrm{d}_i = (\mathrm{d}_{i,\mathrm{img}}, \mathrm{d}_{i,\mathrm{txt}}) \in \mathcal{D}_{\mathrm{cand}}$ of one query $\mathrm{q} = (\mathrm{q}_\mathrm{img}, \mathrm{q}_\mathrm{txt})$, the pointwise reranker takes the concatenation of the query--document pair as input and produces a binary relevance judgment. The input sequence is constructed as:
\begin{equation}
\mathcal{E}_i = \texttt{Concat}\!\left(\, \mathcal{T}_{\mathrm{inst}},\; \mathrm{q}_\mathrm{txt},\; \mathrm{q}_\mathrm{img},\; \mathrm{d}_{i,\mathrm{txt}},\; \mathrm{d}_{i,\mathrm{img}} \,\right),
\end{equation}
where $\mathcal{T}_{\mathrm{inst}}$ is a task-specific instruction prompt. The model is then asked to generate a single token from the binary vocabulary $\{\texttt{Yes}, \texttt{No}\}$, indicating whether $\mathrm{d}_i$ is relevant to $\mathrm{q}$. The relevance score $\mathrm{s}_i$ is derived from the generation probability of the positive token:
\begin{equation}
\mathrm{s}_i = P\!\left(\, \texttt{Yes} \mid \mathcal{E}_i;\, \Phi_{\mathrm{rank}}^{\mathrm{point}} \,\right), \quad \forall\, \mathrm{d}_i \in \mathcal{D}_{\mathrm{cand}}.
\end{equation}
The final candidate ranking follow the descending order of $\mathrm{s}_i$. The prompt template is shown below.

\begin{promptbox}{Prompt Template for Pointwise Reranker $\Phi_{\mathrm{rank}}^{\mathrm{point}}$}
    
    \token{im\_start} system \\
    You are a relevance assessor. Given a query and a candidate document, each consisting of an image and a text description, determine whether the document is relevant to the query. Answer with \texttt{Yes} or \texttt{No}. \\
    \token{im\_end}
    
    \medskip
    
    \token{im\_start} user \\
    Query Text: \{\small$\mathrm{q}_\mathrm{txt}$\} \\
    Query Image: \texttt{<image: $\mathrm{q}_\mathrm{img}$>} \\
    Document Text: \{\small$\mathrm{d}_{i,\mathrm{txt}}$\} \\
    Document Image: \texttt{<image: $\mathrm{d}_{i,\mathrm{img}}$>} \\
    \token{im\_end}
\end{promptbox}

\vspace{0.1cm}
\noindent \underline{\textit{Training Objective}}. The model is optimized via next-token prediction to align the predicted relevance score $\mathrm{s}_i$ with the ground-truth supervision. Formally, let $\mathrm{y}_i \in \{0, 1\}$ be a binary indicator where $\mathrm{y}_i = 1$ if the query--document pair is labeled as relevant ($\text{Yes}$), and $\mathrm{y}_i = 0$ otherwise, the training objective is to minimize the binary cross-entropy loss over a batch $\mathcal{B}$: 
\begin{equation}
\mathcal{L}_{\mathrm{rank}}^{\mathrm{point}} = -\frac{1}{|\mathcal{B}|} \sum_{i \in \mathcal{B}} \Bigl( \mathrm{y}_i \log \mathrm{s}_i + (1 - \mathrm{y}_i) \log(1 - \mathrm{s}_i) \Bigr).
\end{equation}
Under the restricted binary vocabulary $\{\text{Yes}, \text{No}\}$, such one objective effectively compels the MLLM to assign higher probabilistic mass to the ground-truth relevance indicator, thereby capturing the fine-grained multi-modal correlation between queries and candidates.

\vspace{0.1cm}
\noindent \underline{\textit{Data Governance}}. 
To train the pointwise reranker $\Phi_{\mathrm{rank}}^{\mathrm{point}}$, we implement a multi-stage data governance pipeline that transitions from the industry-specific adaptation to reasoning-based label synthesis. (1) Industrial Adaptation: We first conduct continuous pre-training on the ultra-scale prototype-ID library to adapt general-purpose MLLM checkpoints. Within this stage, documents sharing the same instance-ID are treated as positive pairs, while those from disparate IDs serve as negatives, ensuring the model captures fine-grained, industry-specific semantic nuances. 
(2) Reasoning-Based Labeling: Building upon a seed set of human-annotated query-document pairs, we construct an offline reasoning model $\Phi_{\mathrm{rank}}^{\mathrm{reason}}$, through a two-step optimization process. First, we employ state-of-the-art MLLMs (\textit{e.g.}, Gemini-3.0) to generate CoT reasoning trajectories~\cite{wei2022chain} for relevance assessment, which serve as targets for supervised fine-tuning to achieve a cold-start initialization. Second, we apply Group Relative Policy Optimization (GRPO)~\cite{guo2025deepseek-r1} to further align the model with human preference, using a rule-based reward function that evaluates the correctness of predicted relevance labels.
(3) Scalable Annotation and Distillation: The optimized $\Phi_{\mathrm{rank}}^{\mathrm{reason}}$ is then deployed to annotate millions of query-document pairs sampled from real-world search logs. To mitigate noise of CoT and label at scale, we utilize large-scale MLLMs (\textit{e.g.}, Qwen3-VL-32B) as an expert judge, retaining only high-confidence samples where reasoning and labels exhibit cross-model consensus. And finally, we obtain the large-scale, high-purity supervised dataset.

However, the pointwise reranking $\Phi_{\mathrm{rank}}^{\mathrm{point}}$ still suffers from inherent bottlenecks in semantic granularity, contextual reasoning, and computational scalability, as detailed below:

\textit{\textbf{Binary Semantic Reductionism}}: The pointwise solution reduces relevance to a binary judgment (relevant vs. irrelevant), producing overly coarse outputs. In practice, relevance exists on a continuous spectrum, from exact matches to weak correlations, and visually similar candidates often differ in subtle but important ways. Binary classification cannot capture these fine-grained distinctions, leaving the model unable to reliably rank highly competitive candidates.

\textit{\textbf{Absence of Comparative Contextual Awareness}}: The pointwise solution evaluates each document in isolation, ignoring the comparative signals among documents. In practice, effective ranking requires contrastive reasoning, as subtle differences in texture, color, or perspective are far easier to resolve when documents are compared side by side. Without such a global context, the model struggles to maintain a coherent relevance standard, leading to inconsistent and fragile rankings.

\textit{\textbf{Prohibitive Computational Redundancy}}: Pointwise is hampered by suboptimal inference efficiency, particularly in real-time deployment. Executing a full forward pass for every query-document pair leads to a linear increase O(N) in computational complexity as the candidate set expands. Moreover, the query is redundantly re-processed for each evaluation, resulting in significant resource waste. This high inference latency and constrained throughput bottleneck the system's ability to support high-concurrency applications, where response time (RT) is as paramount as ranking precision.

\subsection{Listwise Reranker $\Phi_{\mathrm{rank}}^{\mathrm{list}}$} \label{subsec:listwise}
To transcend the constraints of isolated evaluation, we next propose a listwise reranking pipeline designed to perform joint reasoning across multiple candidates within a shared context. However, a naive listwise idea that ingests the entire hundred-scale candidate set into a single prompt is practically infeasible. Not only does the prefill complexity of Transformer backbone escalate quadratically with the sequence length, but the sheer density of information also risks ``contextual dilution'', where existing MLLMs struggles to maintain fine-grained discriminative resolution over excessive inputs.

To strike a balance between contextual depth and computational tractability, we introduce \textbf{Pailitao-VL-Reranker-List} $\Phi_{\mathrm{rank}}^{\mathrm{list}}$, which decomposes the global ranking problem into a hierarchical ``compare-and-calibrate'' process. To pursue high-precision relevance, we expand the judgment from a coarse binary scale (\texttt{Yes}/\texttt{No}) to the nuanced four-level hierarchy ($0$: instance-level match, $1$: concept-level match, $2$: functional-level match, $3$: irrelevant). By capturing  nuanced relevance levels, the model gains the discriminative power to separate competitive candidates, ensuring reranking is both precise and interpretable.
$\Phi_{\mathrm{rank}}^{\mathrm{list}}$ consists of three synergistic components:

\textit{Chunkwise Local Ranking} (\S\ref{subsubsec:chunkwise}) partitions the candidate list into manageable groups to enable side-by-side comparison among co-occurring documents. By leveraging fine-grained multi-level relevance labels to construct pairwise ranking supervision across documents within each chunk, it captures the subtle hierarchy of relevance that isolated binary evaluation typically overlooks.

\textit{Absolute Relevance Scoring} (\S\ref{subsubsec:absolute}) estimates each candidate's relevance on a fixed, four-level scale, to ensure global comparability across independently processed chunks. Absolute metric serves as a calibrated anchor for a chunk-agnostic reference, enabling consistent cross-chunk score alignment.

\textit{Hybrid Ranking Policy} (\S\ref{subsubsec:hybrid}): Finally, we merge the intra-chunk relative orderings with the absolute scores, producing a coherent global ranking that inherits both the comparative precision of listwise reasoning within each chunk and cross-chunk consistency provided by absolute relevance calibration.

\subsubsection{Chunkwise Local Ranking} \label{subsubsec:chunkwise}
\noindent \underline{\textit{Input/Output Formulation}}.
Given the candidate set $\mathcal{D}_{\mathrm{cand}} = \{\mathrm{d}_1, \dots, \mathrm{d}_N\}$ for a query $\mathrm{q} = (\mathrm{q}_\mathrm{img}, \mathrm{q}_\mathrm{txt})$, we partition it into $K = \lceil N/M \rceil$ chunks $\{\mathcal{C}^{1}, \dots, \mathcal{C}^{K}\}$. For the generality, let $\mathcal{C} = \{\mathrm{d}_1, \dots, \mathrm{d}_m\}$ denote a representative chunk, where $m \leq M$ is the chunk size ({\em e.g.}, $M = 10$). For each chunk, the input sequence $\mathcal{E}$ is constructed by interleaving the query $\mathrm{q}$ with all $m$ candidate documents:
\begin{equation}
\mathcal{E} = \texttt{Concat}\!\left(\, \mathcal{T}_{\mathrm{inst}},\; \mathrm{q}_\mathrm{txt},\; \mathrm{q}_\mathrm{img},\; \mathrm{d}_{1,\mathrm{txt}},\; \mathrm{d}_{1,\mathrm{img}},\; \dots,\; \mathrm{d}_{m,\mathrm{txt}},\; \mathrm{d}_{m,\mathrm{img}} \,\right),
\end{equation}
where $\mathcal{T}_{\mathrm{inst}}$ is a task-specific instruction prompt that assigns each document $\mathrm{d}_i$ a unique index identifier for $i \in \{1, \dots, m\}$.
The model is prompted to select the index of the most relevant document. Rather than relying solely on the generated output, we extract the logits at the generated token position corresponding to each of the $m$ index identifiers as well as an additional \texttt{[NULL]} token (indicating the absence of any relevant document). The logit assigned to each index token serves as the relevance score $\mathrm{s}_i$ for document $\mathrm{d}_i$: a higher logit indicates greater model confidence that $\mathrm{d}_i$ is the most relevant document, and documents are ranked accordingly. This yields a local score vector:
\begin{equation}
\mathbf{s} = \left(\mathrm{s}_1, \dots, \mathrm{s}_{m}, \mathrm{s}_{\texttt{null}}\right) \in \mathbb{R}^{m + 1}.
\end{equation}
By processing all $K$ chunks independently and in parallel, this chunkwise paradigm achieves significant throughput gains over the O(N) pointwise evaluation. The prompt template is shown below.

\begin{promptbox}{Prompt Template for Listwise Reranker $\Phi_{\mathrm{rank}}^{\mathrm{List}}$}
    
    \token{im\_start} system \\
    You are a relevance assessor. Given a query and a list of candidate documents, each consisting of an image and a text description, select the index of the document that best matches the query. If none is relevant, output NULL. \\
    \token{im\_end}
    
    \medskip
    
    \token{im\_start} user \\
    Query Text: \{\small$\mathrm{q}_\mathrm{txt}$\} \\
    Query Image: \texttt{<image: $\mathrm{q}_\mathrm{img}$>} \\    
    Document Text: \{\small$\mathrm{d}_{1,\mathrm{txt}}$\} ~~ Document Image: \texttt{<image: $\mathrm{d}_{1,\mathrm{img}}$>} \\ \relax
    Document Text: \{\small$\mathrm{d}_{2,\mathrm{txt}}$\} ~~ Document Image: \texttt{<image: $\mathrm{d}_{2,\mathrm{img}}$>} \\ \relax
    $\vdots$ \\ \relax
    Document Text: \{\small$\mathrm{d}_{m,\mathrm{txt}}$\} ~~ Document Image: \texttt{<image: $\mathrm{d}_{m,\mathrm{img}}$>} \\
    \token{im\_end}
\end{promptbox}

\vspace{0.1cm}
\noindent \underline{\textit{Training Objective}}.
For each document in one chunk, we construct the multi-level relevance label $\mathrm{y}_i \in \{0, 1, 2, 3\}$, where smaller values indicate higher relevance ($0$: instance-level match, $1$: concept-level match, $2$: functional-level match, $3$: irrelevant). The index and \texttt{[NULL]} tokens are supervised with a chunkwise training objective consisting of two components.

\vspace{0.1cm}
\textit{\textbf{Ranking Loss}}. 
For one given chunk, we define the set of discordant pairs as $\mathcal{P} = \{(i,j) \mid \mathrm{y}_i > \mathrm{y}_j \}$, where document $j$ is more relevant than document $i$. We optimize the following pairwise loss to encourage $\mathrm{s}_j$ to exceed $\mathrm{s}_i$: 
\begin{equation}
\mathcal{L}_{\text{rank}}^{\mathrm{pair}} = \frac{1}{|\mathcal{P}|}
\sum_{(i,j) \in \mathcal{P}}
w_{ij} \cdot \log\left(1 + \exp\left(-\frac{\mathrm{s}_j - \mathrm{s}_i}{\tau}\right)\right),
\end{equation}
where $\tau > 0$ is a temperature hyperparameter, and $w_{ij} = \max\!\bigl(\mathrm{y}_i - \mathrm{y}_j, 1\bigr)$ assigns stronger gradients to pairs with larger relevance gaps.

\vspace{0.1cm}
\textit{\textbf{Null-Token Loss}}.
The \texttt{[NULL]} token serves as a decision boundary to separate instance-level matches ($\mathrm{y} = 0$) from other cases ($\mathrm{y} > 0$). Here, let $\mathcal{S} = \{i \mid \mathrm{y}_i = 0 \}$ be the set of instance-level matches and $\mathcal{O} = \{i \mid \mathrm{y}_i > 0 \}$
be the other set. We enforce that instance-level matches score above $\mathrm{s}_{\texttt{null}}$ while others score below it:
\begin{equation}
\mathcal{L}^{\text{null}}_{\mathrm{rank}} = 
\frac{1}{|\mathcal{S}|} \sum_{i \in \mathcal{S}} \log\left(1 + \exp\left(-\frac{\mathrm{s}_i - \mathrm{s}_{\texttt{null}}}{\tau}\right)\right)
+
\frac{1}{|\mathcal{O}|} \sum_{i \in \mathcal{O}} \log\left(1 + \exp\left(-\frac{\mathrm{s}_{\texttt{null}} - \mathrm{s}_i}{\tau}\right)\right).
\end{equation}

The total chunkwise loss is computed by averaging the components across all $K$ chunks in a batch:
\begin{equation}
\mathcal{L}_{\mathrm{rank}}^{\text{chunk}} = \frac{1}{K}\sum_{k=1}^K \left( \mathcal{L}_{\text{rank}}^{k,\mathrm{pair}} + \mathcal{L}_{\mathrm{rank}}^{k,\text{null}} \right).
\end{equation}

\subsubsection{Absolute Relevance Scoring}
\label{subsubsec:absolute}
While chunkwise ranking enables efficient local ordering, the score $\mathrm{s}$ is only meaningful within a chunk due to the local logit normalization. Scores across different chunks are not directly comparable, leading to inconsistent global ranking. To tackle this, we employ a complementary absolute relevance scoring that estimates the relevance of each query--document pair on one fixed, globally shared scale.

\vspace{0.1cm}
\noindent \underline{\textit{Architecture}}.
The absolute scorer reuses the hidden embeddings already computed during chunkwise inference, requiring \emph{no additional forward pass}. For one chunkwise input sequence $\mathcal{E}$, we extract per-image embeddings directly from the MLLM's final hidden states, by identifying image token positions and grouping them into $m+1$ contiguous segments $\{\mathcal{I}_0, \mathcal{I}_1, \dots, \mathcal{I}_m \}$, where $\mathcal{I}_0$ corresponds to the query and $\mathcal{I}_i$ to the $i$-th candidate document. For the MLLM's final hidden states of each segment, we apply mean pooling over the token length, and then obtain $\mathbf{z} \in \mathbb{R}^{D}$, where $D$ refers to the embedding dimension. 
After that, for each candidate document $i \in \{1,\dots,m\}$, we form a joint representation by concatenating the query embedding $\mathbf{z}_0$ with the document embedding $\mathbf{z}_i$, and pass it through a lightweight MLP head $\Phi_\mathrm{MLP}$ to predict relevance logits $\mathrm{p}_i$: 
\begin{equation}
\mathrm{p}_i = \Phi_\mathrm{MLP} \left( \left[ \mathbf{z}_0 \;\middle\|\; \mathbf{z}_i \right] \right) \in \mathbb{R}^{4}, \quad   \mathrm{r}_i = \mathrm{softmax}(\mathrm{p}_i), 
\end{equation}
where $4$ is the number of relevance levels ($\mathrm{y} \in \{0,1,2,3\}$). At inference, the absolute relevance score $\mathrm{r}_i$ is the probability of the most relevant level. 
Since $\mathrm{r}_i$ is derived from a fixed label taxonomy rather than intra-chunk competition, it provides a calibrated metric that is comparable across all chunks.

\vspace{0.1cm}
\noindent \underline{\textit{Training Objective}}.
The absolute scorer is optimized via a cross-entropy loss over the ground-truth relevance levels $\mathrm{y}_i \in \{0,1,2,3\}$. Let $\mathrm{r}_{i, \mathrm{y}_i}$ denote the predicted probability of the target level $\mathrm{y}_i$ for candidate document $i$, this loss is formulated as:
\begin{equation}
\mathcal{L}^{\text{abs}}_{\mathrm{rank}} = -\frac{1}{|\mathcal{B}|} \sum_{i \in \mathcal{B}} \log \mathrm{r}_{i, \mathrm{y}_i},
\end{equation}
where $\mathcal{B}$ refers to the collection of all candidate documents within the training batch. And the final multi-task objective combines all components:
\begin{equation}
\mathcal{L}_{\mathrm{rank}} = \mathcal{L}^{\text{ntp}}_{\mathrm{rank}} + \mathcal{L}^{\text{chunk}}_{\mathrm{rank}} + \mathcal{L}^{\text{abs}}_{\mathrm{rank}},
\end{equation}
where $\mathcal{L}^{\text{ntp}}_{\mathrm{rank}}$ is the standard next-token prediction loss. This design introduces negligible overhead while ensuring robust cross-chunk comparability.

\subsubsection{Hybrid Ranking Policy}
\label{subsubsec:hybrid}
Chunkwise scorings and absolute scorings provide the complementary signals: (1) \textit{chunkwise scores} $\mathrm{s}^{k}_i$ capture fine-grained relative preferences within each chunk, while (2) \textit{absolute scores} $r^{k}_i$ provide chunk-agnostic absolute relevance estimates. To exploit both local comparative reasoning and global comparability, we adopt a two-stage hybrid ranking policy.

\vspace{0.1cm}
\noindent \underline{\textit{Stage 1: Within-Chunk Ordering}}.
For each chunk $\mathcal{C}^{k}$ ($k \in \{1,\dots,K\})$, we first sort its candidates in descending order of their local chunkwise scores. This yields a locally permuted sequence:
\begin{equation}
\mathrm{d}^{k}_1, \mathrm{d}^{k}_2, \dots, \mathrm{d}^{k}_m
\quad\text{s.t.} \quad
s^{k}_1 \ge s^{k}_2 \ge \cdots \ge s^{k}_m.
\end{equation}

\vspace{0.1cm}
\noindent \underline{\textit{Stage 2: Cross-Chunk Merging}}.
We aggregate the $K$ locally-sorted chunks into one unified global ranking using a multi-way merge procedure. This process ensures that the local relative order within each chunk is strictly preserved. Specifically, we maintain a pointer $\mathrm{t}_{k}$ for the chunk $k$, initially pointing to the top-ranked candidate. In each step, we examine the candidates currently indexed by $\{\mathrm{t}_{k}\}_{k=1}^K$ across all chunks and select the one with the highest absolute score $\mathrm{r}$: 
\begin{equation}
k^\star = \arg\max_{k: \mathrm{t}_k \le m} r^{k}_{\mathrm{t}_k},
\end{equation}
where $k^\star$ identifies the winning chunk. The document $\mathrm{d}^{k^\star}_{\mathrm{t}_{k^\star}}$ is then moved to the global list, and its chunk pointer $\mathrm{t}_{k^\star}$ is advanced to the next position. This greedy selection continues until all candidate documents are merged. This hybrid policy preserves the fine-grained relative orderings established by listwise comparison within each chunk, while leveraging the calibrated absolute scorer to resolve the "global" priority across chunk boundaries. Consequently, the final ranking inherits the benefits of both precise intra-chunk discrimination and robust, chunk-independent estimation.

\section{Experiments} \label{exp}
\subsection{Experimental Settings}

\vspace{0.1cm}
\noindent \textbf{Model Initialization.} We build upon advanced MLLM foundations: \texttt{TBstars-VL-Embedding-3B} is utilized for MLLM2vec of the embedding model, while \texttt{Qwen3-VL-2B}~\citep{Qwen3-VL} is adopted for the reranking model (both point and list). 2B/3B parameter scale is deliberately selected to strike an optimal balance between representational capacity and inference efficiency.

\vspace{0.1cm}
\noindent \textbf{Evaluation Datasets}.
For embedding $\Phi_{\mathrm{embd}}$, the retrieval test set comprises 5K e-commerce queries and 1.5 million e-commerce documents. The annotators assess each document’s relevance to its query, labeling relevance as instance-level matching, concept-level or irrelevant. 

For reranker $\Phi_{\mathrm{rank}}$, the candidate test set comprises 2K e-commerce queries, averaging 50 associated documents per query. Each query--document pair is manually labeled with four types of relevance, {\em i.e.}, instance-level matching (0), concept-level matching (1), functional-level matching (2), irrelevant (3).

All queries here are sourced from real-world user exposure requests, which are subject to various types of noise. From the visual view, noise is low lighting, cluttered backgrounds, physical overlays, and low resolution; while from the textual view, noise is singular/plural mismatches, punctuation inconsistencies, typos, misspellings, and informal abbreviations ({\em e.g.}, ``BBQ'' for ``Barbecue'').

\vspace{0.1cm}
\noindent\textbf{Metrics}. 
To fully meet needs of real-world applications, we use two types of elasticity for evaluation: concept-/category-level relevance ({\em e.g.}, differentiating a sedan from an SUV), instance-level relevance ({\em e.g.}, differentiating a facelifted model from its predecessor based on subtle headlight contours). 

For embedding $\Phi_{\mathrm{embd}}$, we employ an adaptive Hit Rate@$K$ (HR@$K$) as metric, to simultaneously evaluate recall and precision. For each query, relevant documents are defined as those labeled with 0 (instance-level match) or 1 (concept-level match). Let G denote the total number of such relevant documents available in the retrieval set. For a given cutoff $K$, the HR@$K$ is defined as the proportion of relevant items successfully ranked within the top min(K,G) positions:
\begin{equation}
\text{I-HR@}K = \frac{\sum_{i=1}^{\min(K, G)} \mathbb{I}(\text{label}_i \in \{0\})}{\min(K, G)}, \quad
\text{C-HR@}K = \frac{\sum_{i=1}^{\min(K, G)} \mathbb{I}(\text{label}_i \in \{0, 1\})}{\min(K, G)},
\end{equation}
where $\mathbb{I}(\cdot)$ is the indicator function. This adaptive normalization ensures that the metric is calibrated against the maximum achievable hits within the specified search window. Unlike the standard Hit Rate or Precision@k which often use a fixed denominator $K$, our HR@k formulation specifically evaluates the model's ranking density, {\em i.e.}, its ability to consolidate all available ground-truth items at the absolute head of the list. Such a metric is particularly vital for high-precision e-commerce scenarios, where the number of relevant products per query may be smaller than the cutoff $K$. By accounting for the scarcity of positive samples, HR@k provides rigorous/fair assessment of the model's effectiveness in satisfying user intent within limited display slots.

In addition to HR@$K$, we further evaluate the reranker $\Phi_{\mathrm{rank}}$ from a relevance classification perspective to assess its intrinsic precision across different semantic granularities. Since reranker estimates a calibrated relevance score for each candidate document, we could examine its ability to distinguish positive instances from negatives at both instance-level and concept-level.

\underline{\textit{Instance-level Alignment}} evaluates the model's capacity to isolate exact SKU match. We define the positive class as documents with label $\{0\}$, while all other labels $\{1, 2, 3\}$ are treated as negatives.

\underline{\textit{Conceptual-level Alignment}} assesses whether the model can identify the same item or product category. Here, the positive class comprises labels $\{0, 1\}$, and the negative class includes $\{2, 3\}$.

For both settings, we compute Precision, Recall, and F1-Score across the full candidate documents. Unlike ranking-based metrics that focus on relative permutations, these classification metrics provide an absolute measure of the model's decision-making quality. This is particularly crucial for industrial downstream applications---such as automated product filtering and quality assurance---where the model must not only rank items correctly but also exhibit a robust capability to reject irrelevant candidates via an absolute decision threshold.

\vspace{0.1cm}
\noindent\textbf{Hyperparameters.} 
For Pailitao-VL-Embedding $\Phi_{\text{embd}}$, we employ the AdamW optimizer with a peak learning rate of $2 \times 10^{-5}$, a cosine decay schedule, and a warmup ratio of 0.05. To provide sufficiently diverse contrastive signals within each batch, we utilize a global batch size of 256 and train the model for one full epoch. During inference, we adopt the 512-dimensional for industrial-scale retrieval, striking an optimal balance between retrieval precision and computational efficiency.
For Pailitao-VL-Reranker, both $\Phi^{\text{point}}_{\text{rank}}$ and  $\Phi^{\text{list}}_{\text{rank}}$ fine-tune all parameters, including the vision encoder, the vision-language adapter, and the LLM, with AdamW using an initial learning rate of $5 \times 10^{-6}$, cosine decay, and a warmup ratio of 0.03. The batch size is 64, and each image is natively resized so that the number of visual tokens does not exceed 80.  The pointwise variant is trained for 2 epochs and receives both the full image and its cropped region as visual inputs, enhancing robustness to cluttered backgrounds and localized object details. The listwise variant is trained for 5 epochs, with the maximum chunk size being $M = 10$ during both training and inference.

\begin{table*}[t]
\centering
\footnotesize
\setlength{\tabcolsep}{8pt}
\caption{\textbf{Offline Performance of Embedding.} Pailitao-VL-Embedding  is far ahead.} 
\vspace{-0.2cm}
\label{tab:offline-embedding}
\begin{tabular}{lcccccc|c}
\toprule
\multirow{2}{*}{\textbf{Method}} & \multicolumn{3}{c}{\textbf{Instance-Level (I-HR@$K$)}} & \multicolumn{3}{c}{\textbf{Conceptual-Level (C-HR@$K$)}} & \multirow{2}{*}{\textbf{Average}} \\
\cmidrule(lr){2-4} \cmidrule(lr){5-7} 
& \textbf{@1} & \textbf{@5} & \textbf{@10} & \textbf{@1} & \textbf{@5} & \textbf{@10} & \\
\midrule
Qwen3-VL-Embedding  & 57.83 & 58.21 & 60.11 & 73.43 & 72.23 & 71.14 & 65.49 \\
CLIP-IDClass-Embedding  & 60.05 & 60.28 & 62.44 & \underline{80.38} & \underline{80.28} & \underline{79.90} & \underline{70.56} \\
TBStars-VL-Embedding  & \underline{60.40} & \underline{60.50} & \underline{63.36} & 78.27 & 77.56 & 77.18 & 69.55 \\
\midrule
\rowcolor{gray!15}
Pailitao-VL-Embedding       & \textbf{64.52} & \textbf{64.89} & \textbf{67.66} & \textbf{82.01} & \textbf{81.72} & \textbf{81.54} & \textbf{73.73} \\
\bottomrule
\end{tabular}
\end{table*}

\begin{table*}[t]
\centering
\footnotesize
\setlength{\tabcolsep}{5pt}
\caption{\textbf{Offline Performance of Reranker.} Pailitao-VL-Reranker demonstrates the great superiority in both instance-level and concept-level retrieval.}
\vspace{-0.2cm}
\label{tab:offline-rec}
\begin{tabular}{lcccccccc|c}
\toprule
\multirow{2}{*}{\textbf{Method}} & \multicolumn{4}{c}{\textbf{Instance-Level (I-HR@$K$)}} & \multicolumn{4}{c}{\textbf{Conceptual-Level (C-HR@$K$)}} & \multirow{2}{*}{\textbf{Average}} \\
\cmidrule(lr){2-5} \cmidrule(lr){6-9}
& \textbf{@1} & \textbf{@5} & \textbf{@10} & \textbf{@15} & \textbf{@1} & \textbf{@5} & \textbf{@10} & \textbf{@15} & \\
\midrule
Qwen3-VL-Embedding  & 50.10 & 47.79 & 46.75 & 46.23 & 90.59 & 88.72 & 87.33 & 86.38 & 67.99 \\
CLIP-IDClass-Embedding           & 54.48 & 52.38 & 51.24 & 50.60 & 91.57 & 89.05 & 87.36 & 86.16 & 70.36 \\
Qwen3-VL-Reranker   & 51.99 & 49.56 & 48.50 & 47.90 & 92.08 & 90.05 & 88.36 & 87.35 & 69.47 \\
\midrule
\rowcolor{gray!15}
Pailitao-VL-Reranker-Point   & \underline{56.45} & \underline{54.54} & \underline{53.30} & \underline{52.74} & \underline{92.93} & \underline{90.49} & \underline{88.71} & \underline{87.51} & \underline{72.21} \\
\rowcolor{gray!15}
Pailitao-VL-Reranker-List       & \textbf{57.92} & \textbf{55.79} & \textbf{54.74} & \textbf{54.20} & \textbf{94.14} & \textbf{92.36} & \textbf{91.05} & \textbf{90.09} & \textbf{73.79} \\
\bottomrule
\end{tabular}
\end{table*}

\subsection{Offline Embedding Performance} \label{sec:offline emb}
We compare against three SOTA competitors, as follows:

\underline{\textit{Qwen3-VL-Embedding}}~\citep{li2026qwen3-vl-emb} represents the leading VLM2vec in general scenarios, driven by superior data governance and advanced optimization paradigms.

\underline{\textit{TBStars-VL-Embedding}} represents the powerful VLM2vec in industry scenarios, performing large-scale continued training on downstream E-commerce data.

\underline{\textit{CLIP-IDClass-Embedding}} follows the CLIP dual-encoder architecture, inherits data governance of Pailitao-VL-Embedding and is optimized by ultra-scale ID-recognition for absolute anchor learning.

Table~\ref{tab:offline-embedding} provides a comprehensive evaluation. A cross-methodological analysis reveals several key insights regarding domain-specific data and optimization paradigms.

First, the comparison between Qwen3-VL-Embedding and TBStars-VL-Embedding highlights the impact of domain-specific pre-training. TBStars consistently outperforms Qwen3-VL across both levels, particularly achieving a higher I-HR@1 (60.40 {\em vs.} 57.83). This performance gap is consistent with expectations, as TBStars is pre-trained on massive, in-domain industry data, allowing it to capture specialized data distribution more effectively than the general-domain foundation model.

Crucially, the results highlight the inherent advantages of ID-recognition paradigm. Unlike Qwen3-VL and TBStars, which rely on the standard contrastive learning, CLIP-IDClass-Embedding is optimized through ultra-scale classification. Notably, despite its smaller parameter, CLIP-IDClass-Embedding (70.56) significantly outperforms the larger, contrastive-based TBStars-VL (69.55). This empirical evidence suggests that ultra-scale ID-recognition serves as a more effective objective for multi-modal embedding; through providing explicit and dense supervisory signals, it better captures fine-grained semantic boundaries compared to the often noise-prone sampling of contrastive pairs.

Finally, Pailitao-VL-Embedding achieves state-of-the-art across all metrics, reaching an Average score of 73.73. By augmenting the MLLM2vec backbone with ultra-scale ID classification, we improve I-HR@1 from 60.40 (TBStars) to 64.52 at the instance-level evaluation, validating that the classification-based training objective further enhances the MLLM's sophisticated visual understanding, enabling it to better resolve challenging intra-ID variations in complex retrieval scenarios.

\subsection{Offline Reranking Performance} \label{sec:offline rank}
We compare against competitors: Two embedding-based models from \S\ref{sec:offline emb}, \underline{\textit{Qwen3-VL-Embedding}} and \underline{\textit{CLIP-IDClass-Embedding}}, are repurposed as reranking baselines by scoring each candidate via cosine similarity with the query embedding. In addition, we include one dedicated reranking baseline: \underline{\textit{Qwen3-VL-Reranker}}~\citep{li2026qwen3-vl-emb} is a leading pointwise solution in general scenarios, driven by superior data governance and advanced optimization paradigms. It scores each query--document pair through $\text{sigmoid}(\text{logit}(\texttt{yes}) - \text{logit}(\texttt{no}))$.

\noindent\textbf{Top-$K$ Hit Rate.} As shown in Table~\ref{tab:offline-rec}, both of our variants outperform all baselines across instance-level and concept-level metrics at every cutoff. Pailitao-VL-Reranker-List achieves the best overall performance, improving I-HR@1 by $+3.44$ over CLIP-IDClass-Embedding and C-HR@1 by $+2.06$ over Qwen3-VL-Reranker. Compared to the pointwise variant, the listwise one provides further gains, confirming the benefit of combining local chunkwise ordering with global absolute scores.

\noindent\textbf{Relevance Classification.} Table~\ref{tab:offline-cls} reports Precision, Recall, and F1-Score under two label groupings. For embedding models, decision thresholds are tuned on the validation set. Embedding-based models suffer from one clear precision–recall imbalance when repurposed as classifiers ({\em e.g.}, Qwen3-VL-Embedding achieves only 37.86 instance-level precision despite 73.05 recall). Among dedicated rerankers, our pointwise variant already surpasses Qwen3-VL-Reranker by +11.67 instance-level F1, and the listwise variant further improves precision (74.28 vs. 63.85) without sacrificing recall, confirming that chunkwise comparison yields better-calibrated scores and tighter decision boundaries. Our listwise model shows best F1-Score on both instance-level (74.10) and concept-level (91.33), proving not only superior ranking but also robust absolute decision quality for industrial deployment.

\begin{table}[t]
\centering
\small
\setlength{\tabcolsep}{6pt}
\caption{\textbf{Offline performance (Precision, Recall, F1-Score)} on the test set. For embedding models, the decision thresholds are selected on the validation set to maximize F1-Score.}
\label{tab:offline-cls}
\begin{tabular}{lcccccc}
\toprule
\multirow{2}{*}{\textbf{Method}} &
\multicolumn{2}{c}{\textbf{Precision}} &
\multicolumn{2}{c}{\textbf{Recall}} &
\multicolumn{2}{c}{\textbf{F1-Score}} \\
\cmidrule(lr){2-3}
\cmidrule(lr){4-5}
\cmidrule(lr){6-7}
& \textbf{Instance} & \textbf{Concept}
& \textbf{Instance} & \textbf{Concept}
& \textbf{Instance} & \textbf{Concept} \\
\midrule
Qwen3-VL-Embedding     & 37.86 & \underline{87.72} & 73.05 & 54.60 & 49.87 & 67.30 \\
CLIP-IDClass-Embedding     & 60.34 & 78.29 & \underline{75.38} & 93.72 & 67.03 & 85.31 \\
Qwen3-VL-Reranker     & 59.75 & 80.36 & 61.65 & \underline{94.09} & 60.68 & 86.66 \\
\midrule
\rowcolor{gray!15}
Pailitao-VL-Reranker-Point     & \underline{63.85} & 82.50    & \textbf{83.46} & \textbf{95.46}    & \underline{72.35} & \underline{88.51} \\
\rowcolor{gray!15}
Pailitao-VL-Reranker-List & \textbf{74.28} & \textbf{88.73} & 73.93 & \underline{94.09} & \textbf{74.10} & \textbf{91.33} \\
\bottomrule
\end{tabular}
\end{table}

\begin{table*}[t]
\centering
\small
\setlength{\tabcolsep}{5pt} 
\caption{\textbf{Ablation of Pailitao-VL-Reranker-List}. All components contribute together to the best.} 
\vspace{-0.2cm}
\label{tab:ablation}
\begin{tabular}{lcccccccc|c}
\toprule
\multirow{2}{*}{\textbf{Method}} & \multicolumn{4}{c}{\textbf{Instance-Level (I-HR@$K$)}} & \multicolumn{4}{c}{\textbf{Conceptual-Level (C-HR@$K$)}} & \multirow{2}{*}{\textbf{Average}} \\
\cmidrule(lr){2-5} \cmidrule(lr){6-9}
& \textbf{@1} & \textbf{@5} & \textbf{@10} & \textbf{@15} & \textbf{@1} & \textbf{@5} & \textbf{@10} & \textbf{@15} & \\
\midrule
Chunkwise Ranking       & 57.46 & 55.27 & 54.82 & 52.66 & 93.74 & 91.23 & 87.68 & 87.53 & 72.30 \\
+ NULL-Token Loss          & \textbf{58.07} & \textbf{55.95} & 54.60 & 54.03 & 93.93 & 91.61 & 88.60 & 88.02 & 73.10 \\
+ Abs Rel Score          & 57.68 & 55.73 & 54.61 & 54.07 & 94.13 & 92.06 & 90.73 & 89.74 & 73.59 \\
+ Hybrid Ranking     & 57.92 & 55.79 & \textbf{54.82} & \textbf{54.20} & \textbf{94.14} & \textbf{92.36} & \textbf{91.05} & \textbf{90.09} & \textbf{73.79} \\
\bottomrule
\label{tab:ab for reranking}
\end{tabular}
\end{table*}

\vspace{0.1cm}
\noindent\textbf{Component for Reranking}. 
In Table~\ref{tab:ab for reranking}, we ablate the key components of Pailitao-VL-Reranker-List by incrementally adding each module. The \emph{Chunkwise Ranking} uses only chunkwise training. Adding the \emph{NULL-Token Loss} to the training objective improves instance-level metrics at small cutoffs ({\em e.g.}, I-HR@1: 57.46 $\rightarrow$ 58.07) and provides consistent gains on concept-level metrics, indicating that an explicit boundary signal for label-0 items helps the model better separate exact ID matches from near-duplicate candidates. Incorporating the \emph{Abs Rel Score} ({\em i.e.}, the absolute relevance head) further boosts concept-level hit rates. Finally, the full \emph{Hybrid Ranking} variant, which combines chunkwise scores with absolute relevance scores, achieves the best overall performance, demonstrating that local relative ranking and global absolute calibration are complementary signals.

\begin{table}[t]
\footnotesize
\centering
\caption{\textbf{Inference Efficiency Comparison of Reranking} between pointwise and chunkwise. Chunkwise demonstrates overwhelming throughput advantages over pointwise.}
\label{tab:efficiency}
\renewcommand{\arraystretch}{1.15}
\begin{tabular}{l c c}
\toprule
\textbf{Metric} & \textbf{Pointwise} & \textbf{Chunkwise} \\
\midrule
\# MLLM Forward Passes & 800 & 100 \\
Total Time (s) & 18.22 & 7.50 \\
\midrule
Latency / Query (ms) $\downarrow$ & 182.25 & \textbf{75.01} \\
Latency / Pair (ms) $\downarrow$ & 22.78 & \textbf{9.38} \\
\midrule
Throughput (query/s) $\uparrow$ & 5.49 & \textbf{13.33} \\
Throughput (pair/s) $\uparrow$ & 43.90 & \textbf{106.65} \\
\bottomrule
\end{tabular}
\vspace{-0.2cm}
\end{table}

\vspace{0.1cm}
\noindent\textbf{Efficiency for Reranker}. 
Although the above performance is impressive, the reranker efficiency has always been a key bottleneck restricting its deployment of applications. This report evolves the paradigm of generative reranking from pointwise to chunkwise, significantly improving efficiency.
As shown in Table~\ref{tab:efficiency}, we evaluate the inference efficiency on a benchmark of 100 queries with 800 query-document pairs using the vLLM engine on a single NVIDIA A800 GPU. By transition from pointwise to chunkwise, the number of MLLM forward passes is drastically reduced from 800 to 100. The pipeline optimization yields a significant efficiency gain: the per-query latency drops from 182.25 ms to 75.01 ms, representing the $2.4 \times$ speedup. Moreover, the system throughput is boosted from 5.49 to 13.33 queries per second. These empirical results demonstrate that our chunkwise solution effectively mitigates the computational overhead of generative models, ensuring high-throughput processing while maintaining superior ranking quality for large-scale industrial retrieval.

\subsection{Online A/B Results}
To evaluate real-world efficacy, we conduct extensive online A/B testing within the high-concurrency environment of the Pailitao e-commerce platform. Benefiting from the unrelenting optimization efforts of the AI infrastructure team, the system achieves strict real-time performance during deployment. Specifically, the inference latency of Pailitao-VL-Embedding is compressed to 67 ms per query, while Pailitao-VL-Reranker-List achieves an average latency of 76 ms per query.

Multi-Day average statistics demonstrate that these technical breakthroughs translated directly into substantial business impact: Pailitao-VL-Embedding yielded the 2\% GMV gain across the entire platform's traffic, while Pailitao-VL-Reranker-List delivered a significant 6\% GMV gain within standardized product categories. Besides that, in some emerging scenarios of AI search (such as e-commerce SKU-price comparison), Pailitao-VL-Reranker-List has brought about a impressive GMV gain of 20\%, demonstrating the strong potentials.

These outcomes validate Pailitao-VL’s substantial business value and the engineering resilience essential for scaling multi-modal retrieval in demanding industrial environments.

\section{Conclusion}
In this work, we presented Pailitao-VL, a comprehensive multi-modal retrieval system engineered for high-precision, real-time industrial search. Our primary contribution lies in two fundamental paradigm shifts. First, we transitioned the embedding paradigm from traditional contrastive learning to an absolute ID-recognition task. Through anchoring instances to a globally consistent latent space defined by billions of semantic prototypes, we successfully overcome the stochasticity and granularity bottlenecks inherent in existing embedding solutions. Second, we evolved the generative reranker from isolated pointwise evaluation to the ``compare-and-calibrate'' listwise policy. By synergizing chunk-based comparative reasoning with calibrated absolute relevance scoring, the system achieves nuanced discriminative resolution while circumventing the prohibitive latency typically associated with conventional reranking methods. Extensive offline benchmarks and online A/B tests on Alibaba’s e-commerce platform confirm that Pailitao-VL achieves state-of-the-art performance and delivers substantial business impact. This work demonstrates a robust and scalable path for deploying advanced MLLM-based retrieval architectures in demanding, large-scale production environments.

\section{Contributors}
\noindent\textbf{Core Contributors}: Lei Chen, Chen Ju, Xu Chen, Zhicheng Wang

\noindent\textbf{Algorithm Contributors}: Yuheng Jiao, Hongfeng Zhan, Zhaoyang Li, Shihao Xu, Lianyu Du, Xiaohan Ye, Yunmeng Shu, Jinsong Lan, Xiaoyong Zhu, Bo Zheng

\noindent\textbf{Engineering Contributors}: Zhixiang Zhao, Tong Jia, Dongzi Zhao, Shuaiqi Jia, Jiacheng Li, Zhilong Hu, Gaofeng Li, Bo Xie, Chong Ma, Tengfei Xie

\noindent\textbf{Special Contributors} (TBStars-VL-Embedding \& Adaptation): Lin Li, Yuan Gao, Jun Song

\clearpage
\bibliographystyle{plainnat}
\bibliography{refs}

\end{document}